\documentclass[11pt,a4paper]{article}
\usepackage{amssymb}
\usepackage{amsmath}
\usepackage{amsfonts}
\usepackage{dsfont}
\usepackage{amsthm}
\usepackage{mathrsfs}
\usepackage{hyperref}
\usepackage{color}
\usepackage[margin=2.41cm]{geometry}
\usepackage[all,cmtip]{xy}
\usepackage[utf8]{inputenc}
\usepackage{graphicx}
\usepackage{varwidth}
\usepackage{comment}

\usepackage{upgreek}
\usepackage{rotating}

\usepackage{tikz}
\usetikzlibrary{shapes.geometric}

\usepackage[shortlabels]{enumitem}

\definecolor{darkred}{rgb}{0.8,0.1,0.1}
\hypersetup{
     colorlinks=false,         
     linkcolor=darkred,
     citecolor=blue,
}

\theoremstyle{plain}

\theoremstyle{definition}

\numberwithin{equation}{section}

\def\nn{\nonumber}

\def\bbA{\mathbb{A}}
\def\bbK{\mathbb{K}}
\def\bbR{\mathbb{R}}

\def\bbZ{\mathbb{Z}}

\def\Hom{\mathrm{Hom}}

\def\End{\mathrm{End}}

\def\Der{\mathrm{Der}}
\def\Sym{\mathrm{Sym}}

\def\id{\mathrm{id}}

\def\dd{\mathrm{d}}

\def\dim{\mathrm{dim}}
\def\1{\mathbf{1}}
\def\oone{\mathds{1}}

\def\2AQFT{\mathbf{2AQFT}}

\def\dR{\mathrm{dR}}

\def\QCoh{\mathbf{QCoh}}

\def\Mod{\mathbf{Mod}}
\def\dgMod{\mathbf{dgMod}}

\def\CDGA{\mathbf{CDGA}}

\def\O{\mathcal{O}}

\def\T{\mathsf{T}}

\def\g{\mathfrak{g}}

\def\spec{\mathrm{Spec}\,}
\def\dCrit{\mathrm{dCrit}}

\def\holim{\mathrm{holim}}
\def\colim{\mathrm{colim}}

\def\BV{\mathrm{BV}}
\def\CE{\mathrm{CE}}

\def\tot{\mathrm{tot}}

\newcommand\und[1]{\underline{#1}}

\DeclareMathOperator*{\Motimes}{\text{\raisebox{0.25ex}{\scalebox{0.8}{$\bigotimes$}}}}

\def\sk{\vspace{1mm}}

\makeatletter
\let\@fnsymbol\@alph
\makeatother

\title{%
Classical BV formalism for group actions
}

\author{%
Marco Benini$^{1,2,a}$,  Pavel Safronov$^{3,b}$\ and\ 
Alexander Schenkel$^{4,c}$\vspace{4mm}\\
{\small ${}^1$ Dipartimento di Matematica, Universit\`a di Genova,}\\
{\small Via Dodecaneso 35, 16146 Genova, Italy.}\vspace{2mm}\\
{\small ${}^2$ INFN, Sezione di Genova,}\\
{\small Via Dodecaneso 33, 16146 Genova, Italy.}\vspace{2mm}\\
{\small ${}^3$ School of Mathematics,  University of Edinburgh, }\\
{\small Edinburgh EH9 3FD,  United Kingdom.}\vspace{2mm}\\
{\small ${}^4$ School of Mathematical Sciences, University of Nottingham,}\\
{\small University Park, Nottingham NG7 2RD, United Kingdom.}\vspace{4mm}\\
{\small \begin{tabular}{ll}
Email: & ${}^a$~\texttt{benini@dima.unige.it}\\
& ${}^b$~\texttt{p.safronov@ed.ac.uk}\\
& ${}^c$~\texttt{alexander.schenkel@nottingham.ac.uk}\vspace{2mm}
\end{tabular}
}
}

\date{September 2021}

\begin{document}

\maketitle

\vspace{-7mm}

\begin{abstract}
\noindent We study the derived critical locus of a function $f:[X/G]\to \mathbb{A}_{\mathbb{K}}^1$ on the quotient stack of a smooth affine scheme $X$ by the action of a smooth affine group scheme $G$.  It is shown that $\mathrm{dCrit}(f) \simeq [Z/G]$ is a derived quotient stack for a derived affine scheme $Z$, whose dg-algebra of functions is described explicitly.  Our results generalize the classical BV formalism in finite dimensions from Lie algebra to group actions.
\end{abstract}

\vspace{-2mm}

\vspace{-2mm}

\vspace{-1mm}

\renewcommand{\baselinestretch}{0.8}\normalsize
\tableofcontents
\renewcommand{\baselinestretch}{1.0}\normalsize



\section{\label{sec:intro}Introduction and summary}
The Batalin-Vilkovisky (BV) formalism \cite{BV} is a time-honored
and powerful method to describe gauge theories
and their quantization.  It plays a central role in modern
developments in mathematical quantum field theory (QFT),
in particular in the factorization algebra approach of Costello and Gwilliam \cite{CostelloGwilliam,CostelloGwilliam2}
and in higher categorical generalizations of algebraic QFT 
\cite{FredenhagenRejzner,FredenhagenRejzner2,BSWhomotopy,BSreview,LinearYM}. 
Furthermore, it provides an excellent framework to study the rich interplay between
gauge theories and boundaries or defects \cite{CMR,CMR2}.
\sk

Despite these immense successes, the traditional BV formalism
has also certain shortcomings.  Most notable is the issue that
it is intrinsically perturbative,  even at the classical level.
This is due to the fact that only the Lie algebra of infinitesimal gauge transformations,
but not the whole gauge group,  enters the description of a
gauge theory in terms of the BV formalism.  There can be drastic differences between 
infinitesimal and finite gauge transformations,  even for the simplest examples
of gauge theories.  For instance,  the two Abelian gauge theories with 
structure groups $\bbR$ and $U(1)$ are indistinguishable at the infinitesimal level,
but not at the global level due to features such
as Aharonov-Bohm phases and magnetic charges (i.e.\ Chern classes) in the case of $U(1)$.
Hence,  the traditional BV formalism is by construction blind 
to such global features.
\sk

A global geometric framework that has a great potential to refine and generalize
the traditional BV formalism is currently developed 
in the context of {\em derived algebraic geometry}.  
We refer the reader to \cite{DAG,DAG2,Pridham}
for the details and also to \cite{Calaque,SafronovLecture} 
for less technical introductions to this subject.  Loosely speaking,
derived algebraic geometry introduces a new  
concept of spaces, the so-called {\em derived stacks},  which when applied to the
context of gauge theories  encode not only the gauge fields, but also the ghosts 
and the antifields.  Furthermore,  many structures and constructions familiar
from the BV formalism acquire a mathematically precise 
interpretation within derived algebraic geometry, e.g.\ the antibracket
may be identified with a shifted symplectic (or Poisson) structure
on a derived stack.
\sk

The most direct way in which derived algebraic geometry relates to the BV formalism 
is via so-called {\em derived critical loci}.  Informally,  a derived critical locus 
is a derived stack that refines the space of critical points of a function $f: Y\to \bbR$,
e.g.\ an action function of a physical system. 
While retaining a simple and direct interpretation in terms of critical points, 
the derived critical locus captures refined information encoded both by the antifields of the BV formalism 
and by the ghost fields. The latter are inherited from the structure of the ``space'' of fields 
$Y$.  For instance,  for theories with infinitesimal gauge symmetries, 
$Y =[X/\g]$ is a stack describing a Lie algebra quotient
and,  in the case one considers the whole gauge group, 
$Y=[X/G]$ is a stack describing a group quotient.  
It is in general difficult to provide an explicit model 
for the derived critical locus of a function $f:Y\to\bbR$. 
There are, however, two instances where this task has been accomplished, 
showing the close ties between derived critical loci and the BV formalism. 
(1)~When $Y$ is an ordinary space, i.e.\ an affine scheme,
an explicit algebraic model for the derived critical locus
is presented in \cite{Vezzosi}. The result matches the 
classical BV formalism for theories without gauge symmetries.
(2)~The case of a formal stack $Y=[X/\g]$ is studied in
\cite{CostelloGwilliam2}, where it is shown that
the derived critical locus matches the 
BV formalism for theories with infinitesimal gauge symmetries.
The technical details for this case have been worked out in a 
seminar by Schreiber \cite{Schreiber}.
\sk

The goal of this paper is to present an explicit model for the derived
critical locus of a function on a quotient stack $Y=[X/G]$,
i.e.\ we are interested in theories with finite gauge symmetries.
This case lies beyond the scope of the traditional BV formalism, 
which is intrinsically infinitesimal, 
and, as explained above,  it is central to investigate global (as opposed to infinitesimal) aspects of gauge theories.
To allow for a rigorous discussion within the framework of derived algebraic geometry,  
we restrict our attention to the case of finite-dimensional $X$ and $G$.
Unfortunately, this excludes the interesting case of field theories.
However, we believe that it should be possible to amend our final results 
with functional analytical techniques similar to those used in the traditional BV formalism
\cite{CostelloGwilliam,CostelloGwilliam2,FredenhagenRejzner,FredenhagenRejzner2}
in order to make them applicable also to field theories.  We plan to address this point
in future works.
\sk

The outline of the remainder of this paper is as follows:
Section \ref{sec:DAG} contains a very brief review of
quotient stacks and derived critical loci in derived algebraic geometry.
It is then shown in Section \ref{sec:reduction} that
the derived critical locus of a function $f$
on a quotient stack $[X/G]$ admits a more explicit and computationally convenient
description in terms of a derived quotient stack $\dCrit(f) \simeq [Z/G]$ for some 
derived affine scheme $Z$. The latter can be described very explicitly
by computing certain homotopy pushouts in the model category
$\CDGA^{\leq 0}$ of non-positively graded commutative dg-algebras, 
which we carry out in detail in Section \ref{sec:explicit}.
Sections \ref{sec:DAG}, \ref{sec:reduction} and \ref{sec:explicit}
use techniques from derived
algebraic geometry and model category theory.  Readers who are less 
familiar with such techniques might prefer to start from Section \ref{sec:functions},
where we describe in very explicit terms the function dg-algebra $\O(\dCrit(f))$
and the quasi-coherent sheaf dg-category $\QCoh(\dCrit(f))$ of the derived critical locus
$\dCrit(f)$. (Some algebraic preliminaries that are
useful for reading Section \ref{sec:functions}
are reviewed in Subsection \ref{subsec:prelims}.)
The $(-1)$-shifted symplectic structure on $\dCrit(f)$
is then described very explicitly in Section \ref{sec:symplectic}.
In the final Section \ref{sec:comparison}, we compare our results
to the traditional BV formalism for Lie algebra actions.  
We observe in particular that there exists a van Est-type map
$\mathrm{vE}: \O(\dCrit(f))\to \O(\BV(f))$ from the function dg-algebra
of the derived critical locus $\dCrit(f)$ for quotient stacks 
to the dg-algebra of the traditional BV formalism for Lie algebra actions.
This map is however in general not a quasi-isomorphism, see e.g.\ \cite{Crainic},
which reflects the differences between infinitesimal and finite gauge transformations.

\paragraph{Notation and conventions:} We fix once and for all a field $\bbK$
of characteristic $0$.  All vector spaces and algebras
appearing in this paper will be over $\bbK$. 
For a ($\bbZ$-)graded vector space $V$ and an integer $k\in\bbZ$,
the $k$-shifted graded vector space $V[k]$ is defined by 
$V[k]^n:= V^{k+n}$. Concerning complexes of vector spaces, 
we work with cohomological degree conventions,
i.e.\ the differentials increase the degree by $+1$.
For a cochain complex $V$ and an integer $k\in\bbZ$,
the $k$-shifted cochain complex $V[k]$ is defined by 
equipping the $k$-shifted underlying graded vector space
with the differential $\dd_{V[k]}:=(-1)^k\,\dd_V$.


\section{\label{sec:DAG}Derived critical locus for quotient stacks}
In this section we recall the definition of the derived critical locus for quotient stacks.
\sk

Let $X$ be a derived stack with a right action of a smooth affine group scheme $G=\spec H$,
i.e.\ $H$ is a finitely generated smooth commutative Hopf algebra.
The corresponding derived quotient stack $[X/G]$ is defined as the colimit
\begin{flalign}\label{eqn:quotient-stack}
[X/G] \,:=\, \colim\Big(
\xymatrix@C=1em{
X ~&~ \ar@<0.5ex>[l] \ar@<-0.5ex>[l] X\times G~&~ \ar@<1ex>[l] \ar[l] \ar@<-1ex>[l] X\times G^2 ~&~ \ar@<0.5ex>[l] \ar@<-0.5ex>[l]\ar@<1.5ex>[l] \ar@<-1.5ex>[l]\cdots
}\Big)
\end{flalign}
in the $\infty$-category of derived stacks
of the simplicial diagram associated with the $G$-action  $X\times G \to X$  on $X$. 
\sk

We are particularly interested in the case where $X = \spec A$ is a smooth affine scheme,
i.e.\ $A$ is a finitely generated smooth commutative algebra.  In this case
$[X/G]$ is an ordinary (i.e.\ non-derived) quotient stack.
Let us consider a function $f : [X/G]\to \bbA^1_\bbK$,
where the target is the affine scheme $\bbA_{\bbK}^1 = \spec\bbK[x]$
representing the line.  Note that such a function on $[X/G]$
is the same datum as a $G$-equivariant function 
$\overline{f}:X\to \bbA_{\bbK}^1$ on $X$, where $\bbA_{\bbK}^1$ 
is equipped with the trivial $G$-action.
\sk

We are now ready to define the derived critical locus of $f$. Applying the de Rham differential, 
we obtain a section
\begin{flalign}
\dd^{\dR}f  \,:\, [X/G]~\longrightarrow~ T^\ast[X/G]
\end{flalign}
of the cotangent bundle $\pi: T^\ast [X/G]\to [X/G]$ over the quotient stack $[X/G]$.  The derived critical locus
of $f$ is then given by intersecting
this section with the zero section $0 : [X/G]\to T^\ast[X/G]$,
i.e.\ it is defined as the fiber product
\begin{flalign}\label{eqn:dCritdiagram}
\xymatrix@R=3em@C=3em{
\ar@{-->}[d]\dCrit(f) \ar@{-->}[r] ~&~[X/G]\ar[d]^-{0}\\
[X/G]\ar[r]_-{\dd^{\dR}f}~&~ T^\ast[X/G]
}
\end{flalign}
in the $\infty$-category of derived stacks.
\sk

The derived critical locus carries a canonical $(-1)$-shifted
symplectic structure.  The relevant argument is as follows, see e.g.\ \cite{Albin} for more details.
The total space $T^\ast[X/G]$ of the 
cotangent bundle carries an exact $0$-shifted symplectic
structure $\omega = \dd^{\dR}\lambda$, where 
$\lambda$ denotes the tautological $1$-form on $T^\ast[X/G]$, cf.\ \cite{CalaqueTangent}. 
Both $\dd^{\dR} f : [X/G]\to T^{\ast}[X/G]$ and 
the zero section $0 : [X/G] \to T^\ast [X/G]$ carry
an exact Lagrangian structure.  The fiber product in \eqref{eqn:dCritdiagram}
is thus an intersection of two Lagrangians in a $0$-shifted symplectic derived stack,
which by general results from derived algebraic geometry 
defines a canonical $(-1)$-shifted symplectic structure on $\dCrit(f)$.


\section{\label{sec:reduction}Reduction to derived affine schemes}
The definition of the derived critical locus given by \eqref{eqn:dCritdiagram} is presented 
by a fiber product of \emph{derived stacks}. The goal of this section is to show that
one can rewrite $\dCrit(f) \simeq [Z/G]$ as a quotient of a \emph{derived affine scheme} $Z$ by $G$.
\sk

To compute the fiber product in \eqref{eqn:dCritdiagram},
we have to find a more explicit 
model for the total space $T^\ast[X/G]$ of  the cotangent bundle.
As shown in \cite{SafronovImplosion},  there exists an equivalence of 0-shifted symplectic derived stacks
\begin{flalign}\label{eqn:symplecticreduction}
T^\ast[X/G] \,\simeq\, [T^\ast X/\!\! /G]
\end{flalign}
between the total space $T^\ast[X/G]$ and the {\em symplectic reduction}
$[T^\ast X/\!\! /G]$ of the Hamiltonian $G$-space $(T^\ast X,\omega,\mu)$
given by 
\begin{itemize}
\item the cotangent bundle $T^\ast X$ over $X$, 
\item its canonical symplectic structure
$\omega = \dd^{\dR}\lambda\in \Omega^2(T^\ast X)$, 
where $\lambda\in \Omega^1(T^\ast X)$ 
denotes the tautological $1$-form on $T^\ast X$,  and
\item the moment map $\mu: T^\ast X\to \g^\ast$ defined by 
$\mu(\xi) := -\iota_{\rho(\xi)}\lambda$, 
for all elements $\xi\in\g$ of the  Lie algebra $\g$ of $G$, 
i.e.\ the contraction of minus the tautological $1$-form $-\lambda$ 
against the vector fields $\rho(\xi)$ associated with the action 
of the Lie algebra $\g$ on $T^\ast X$.
\end{itemize}
We refer the reader to Section \ref{sec:explicit} for a more
explicit and detailed description of this Hamiltonian $G$-space.
Recall that symplectic reduction $[T^\ast X/\!\!/ G]$ 
is a two-step construction. First, one takes the fiber product
\begin{flalign}\label{eqn:mu-10} 
\xymatrix@R=3em@C=3em{
\ar@{-->}[d]\mu^{-1}(0) \ar@{-->}[r]~&~ \mathrm{pt}\ar[d]^-{0}\\
T^\ast X \ar[r]_-{\mu}~&~ \g^\ast
}
\end{flalign}
in the $\infty$-category of derived stacks, which determines
the derived zero locus of the moment map $\mu$,  and then one forms
the derived quotient stack 
\begin{flalign}\label{eqn:symplectic-red} 
[T^\ast X/\!\!/G]\,:=\, [\mu^{-1}(0)/G]
\end{flalign} 
defined according to \eqref{eqn:quotient-stack} 
by means of the inherited $G$-action on $\mu^{-1}(0)$.
Using this description and the equivalence in \eqref{eqn:symplecticreduction},
one finds that the derived critical locus $\dCrit(f)$ 
introduced in \eqref{eqn:dCritdiagram} is equivalent to 
the fiber product
\begin{flalign}\label{eqn:dCritquotientfiberproduct} 
\xymatrix@R=3em@C=3em{
\dCrit(f) \ar@{-->}[r] \ar@{-->}[d] &  [X/G] \ar[d]^-{0} \\
[X/G] \ar[r]_-{\dd^{\dR}f} & [\mu^{-1}(0)/G]
}
\end{flalign}
in the $\infty$-category of derived stacks.
\sk

As explained in \eqref{eqn:dCriteasyproof1} and \eqref{eqn:dCriteasyproof2} below, 
the derived critical locus may therefore be 
described as a derived quotient stack
\begin{flalign}\label{eqn:dCriteasy}
\dCrit(f)\,\simeq\, [Z/G]\quad,
\end{flalign}
where $Z$ and its inherited $G$-action are 
defined by the fiber product
\begin{flalign}\label{eqn:dCriteasydiagram}
\xymatrix@R=3em@C=3em{
\ar@{-->}[d]Z \ar@{-->}[r] ~&~X\ar[d]^-{0}\\
X\ar[r]_-{\dd^{\dR} f}~&~ \mu^{-1}(0)
}
\end{flalign}
in the $\infty$-category of derived stacks.
This alternative description is 
very useful for the following reason: 
As a consequence of our assumption that $X=\spec A$ is a smooth affine scheme, 
both the derived stacks $\mu^{-1}(0)$ defined in \eqref{eqn:mu-10} and $Z$ defined in 
\eqref{eqn:dCriteasydiagram} are derived affine schemes, i.e.\
they admit purely algebraic models in terms of commutative dg-algebras in non-positive degrees. 
These facts make the description of the derived critical locus in 
\eqref{eqn:dCriteasy} particularly explicit and computationally convenient. 
\sk

The relevant argument to prove the equivalence in \eqref{eqn:dCriteasy} is as follows:
By definition,  $\mu^{-1}(0)\to [\mu^{-1}(0)/G]$ is an effective epimorphism. 
Since effective epimorphisms are stable under pullback,  the projection map
\begin{flalign}\label{eqn:dCriteasyproof1}
\nn X\times_{\mu^{-1}(0)} X&\, \simeq\, ([X/G]\times [X/G])\times_{[\mu^{-1}(0)/G]\times [\mu^{-1}(0)/G]} \mu^{-1}(0)\\
\nn &\,\to\,([X/G]\times [X/G])\times_{[\mu^{-1}(0)/G]\times [\mu^{-1}(0)/G]} [\mu^{-1}(0)/G]\\
&\,\simeq\,[X/G]\times_{[\mu^{-1}(0)/G]} [X/G]
\end{flalign}
is also an effective epimorphism. Therefore,
\begin{flalign}\label{eqn:dCriteasyproof2}
[Z/G] \,=\,  [(X\times_{\mu^{-1}(0)} X)/G]\, \simeq\, [X/G]\times_{[\mu^{-1}(0)/G]} [X/G]\, \simeq\, \dCrit(f)\quad,
\end{flalign}
where the first step uses the definition of $Z$ in \eqref{eqn:dCriteasydiagram}
and the last step uses the description of the derived critical locus 
in \eqref{eqn:dCritquotientfiberproduct}. 
\sk

Let us also provide a more conceptual viewpoint on the presentation 
\eqref{eqn:dCriteasy} of the derived critical locus of $f:[X/G]\to \bbA_{\bbK}^1$. 
We may identify the fiber product defining $Z$ in \eqref{eqn:dCriteasydiagram} with the fiber product
\begin{flalign}\label{eqn:dCritBVreduction}
\xymatrix@R=3em@C=3em{
\ar@{-->}[d]Z \ar@{-->}[r] ~&~\mathrm{pt}\ar[d]^-{0}\ar[d]\\
\dCrit(\overline{f})\ar[r]_-{\overline{\mu}}~&~ \g^\ast[-1]
}
\end{flalign}
where $\dCrit(\overline{f})$ is the derived critical locus of the underlying 
$G$-equivariant function $\overline{f}:X\to\bbA_{\bbK}^1$. 
In other words, we have a $(-1)$-shifted moment map $\overline{\mu}:\dCrit(\overline{f})\to \g^\ast[-1]$ 
and the derived critical locus $\dCrit(f)$ in the presentation \eqref{eqn:dCriteasy}
is the $(-1)$-shifted symplectic reduction of $\dCrit(\overline{f})$. This point of view is developed further in \cite{AnelCalaque}.


\section{\label{sec:explicit}Explicit model}
The aim of this section is to compute explicitly 
a derived affine scheme $Z$ that models the fiber product in 
\eqref{eqn:dCriteasydiagram}. 
This is a purely algebraic problem that can be addressed by 
working in the model category $\CDGA^{\leq 0}$ of non-positively 
graded commutative dg-algebras. The associated
derived quotient stack $[Z/G]\simeq \dCrit(f)$
provides a model for the derived critical locus, cf.\ \eqref{eqn:dCriteasy}.

\subsection{\label{subsec:prelims}Preliminaries on commutative algebras and Hopf algebras}
Because $X=\spec A$ is by our hypotheses smooth and affine,  the concepts 
of differential forms and vector fields on $X$ admit a purely algebraic description 
that we shall briefly recall.  Let $A$ be an arbitrary finitely generated smooth commutative algebra.
Given any $A$-module $M$,  the $A$-module $\Der(A,M)$ of {\em derivations} 
on $A$ with values in $M$ consists of all linear maps
$D : A\to M$ that satisfy the Leibniz rule
$D(a\,a^\prime) = D(a)\,a^\prime + a\,D(a^\prime)$, for all $a,a^\prime \in A$.
The functor $\Der(A,-) : \Mod_A \to \Mod_A$ is co-representable
by an $A$-module $\Omega^1_A$ that is called the module of {\em K\"ahler $1$-forms} on $A$,
i.e.\ there is a natural isomorphism $\Der(A,-)\cong \Hom^{}_A(\Omega^1_A,-)$.
From this natural isomorphism one obtains a universal derivation
$\dd^{\dR} : A\to\Omega^1_A$ that is called the {\em de Rham differential}.
The de Rham dg-algebra $\Omega^\bullet_A :=  \bigwedge_A^\bullet \Omega^1_A$
is defined as usual by forming the exterior algebra over $A$ of the $A$-module $\Omega^1_A$
and extending $\dd^\dR$ via the graded Leibniz rule to a differential of degree $+1$.
\sk

Because $A$ was assumed to be finitely generated and smooth, 
it follows that the $A$-module $\Omega^1_A$ of K\"ahler $1$-forms 
is finitely generated and projective,  hence it is dualizable in the symmetric 
monoidal category $(\Mod_A,\otimes_A^{},A)$.  We denote its dual by
\begin{flalign}
\T_{\! A}\,:=\, \Hom^{}_A(\Omega^1_A,A) \,\cong\, \Der(A,A)
\end{flalign}
and the associated evaluation and coevalation morphisms by
\begin{flalign}
\mathrm{ev}\,:\, \Omega^1_A\otimes_A^{} \T_{\! A} ~\longrightarrow~ A \quad,\qquad \mathrm{coev} \,:\, A~\longrightarrow~\T_{\! A}\otimes_A^{} \Omega^1_A\quad.
\end{flalign}
The $A$-module $\T_{\! A}$ should be interpreted as the module of vector fields
and $\mathrm{ev}$ is the duality pairing between $1$-forms and vector fields.
\sk

In the case where $G = \spec H$ is a smooth affine group scheme, 
the commutative algebra $H$ carries further the structure of a Hopf algebra,
i.e.\ it is endowed with a coproduct $\Delta :H\to H\otimes H$, a counit
$\epsilon : H\to\bbK$ and an antipode $S : H\to H$. 
We shall use the standard Sweedler notation
\begin{subequations}
\begin{flalign}
\Delta h = h_{\und{1}}\otimes h_{\und{2}}\qquad\text{(summation understood)}
\end{flalign}
to denote the coproduct of $h\in H$, and more generally
\begin{flalign}
\Delta^n h =  h_{\und{1}}\otimes\cdots\otimes h_{\und{n+1}}\qquad\text{(summation understood)}
\end{flalign}
\end{subequations}
for iterated applications of the coproduct. (Note that,  due to coassociativity of the coproduct,
it makes sense to write $\Delta^2 := (\Delta\otimes\id)\Delta= (\id\otimes\Delta)\Delta$
for the two-fold application of $\Delta$, and similarly $\Delta^n$ 
for the $n$-fold application.) From this algebraic 
perspective, the dual Lie algebra of $G=\spec H$ is defined as the quotient
\begin{flalign}
\g^\ast\,:=\, \frac{H^+}{ H^{+\,2}}
\end{flalign}
of the augmentation ideal $H^+ := \ker(\epsilon : H\to \bbK)\subseteq H$ by its square.
The coadjoint $H$-coaction on $\g^\ast$ is induced from the right adjoint coaction
$\delta: H\to H \otimes H\,,~h\mapsto h_{\und{2}}\otimes S(h_{\und{1}})\,h_{\und{3}} $
of the Hopf algebra $H$ on itself, where $S$ denotes the antipode.  The Lie algebra $\g$ of $G=\spec H$
is defined as the dual of $\g^\ast$ and it is given explicitly by 
\begin{flalign}
\g\, :=\, \Hom_\bbK^{}(\g^\ast,\mathbb{K})\, \cong\, \Der_\epsilon(H,\bbK)\quad.
\end{flalign}
The latter is the vector space of derivations relative to the counit $\epsilon:H\to\bbK$, i.e.\
linear maps $\xi : H\to \bbK$ satisfying $\xi(h\,h^\prime) = \xi(h)\,\epsilon(h^\prime) + \epsilon(h)\,\xi(h^\prime)$, 
for all $h,h^\prime\in H$. The Lie bracket $[\xi,\xi^\prime] $, for $\xi,\xi^\prime\in \g$,  is
the relative derivation $[\xi,\xi^\prime]: H\to \bbK$ defined by
\begin{flalign}
[\xi,\xi^\prime](h)\,:=\, \xi(h_{\und{1}})\,\xi^\prime(h_{\und{2}}) - \xi^\prime(h_{\und{1}})\,\xi(h_{\und{2}})\quad,
\end{flalign}
for all $h\in H$.
\sk

A right action $X\times G\to X$ of $G=\spec H$ on $X=\spec A$
is given algebraically by endowing $A$ with the structure
of a right $H$-comodule algebra.  We shall use again
the Sweedler notation 
\begin{flalign}
\delta a \,:= \, a_{\und{0}}\otimes a_{\und{1}}\qquad\text{(summation understood)}
\end{flalign}
to denote the right coaction $\delta :A\to A\otimes H $ on $a\in A$.
From this right coaction one obtains a Lie algebra action
$\rho : \g \to \T_{\! A}\cong \Der(A,A)$  via the formula
\begin{flalign}\label{eqn:Liealgebraaction}
\rho(\xi)(a) \,:=\, a_{\und{0}}\, \xi(a_{\und{1}})\quad,
\end{flalign}
for all $\xi\in\g$ and $a\in A$.  The right $H$-coaction
$\delta : A\to A\otimes H$ induces to the $A$-module
$\Omega^1_A$ of K\"ahler $1$-forms and consequently also 
to its dual $A$-module $\T_{\! A}$ of vector fields,  endowing them with the structure
of right $H$-comodule $A$-modules.  Explicitly, the $H$-coaction
on K\"ahler $1$-forms is given by describing $\Omega^1_A= I/I^2$
as the quotient of the ideal $I:=\ker(m:A\otimes A \to A)\subseteq A\otimes A$ 
determined by the kernel of the multiplication map by its square and using
the tensor product coaction on $A\otimes A$, i.e.\ $\delta(a\otimes a^\prime) = 
a_{\und{0}}\otimes a^\prime_{\und{0}}\otimes a_{\und{1}}\,a^\prime_{\und{1}}$,
for all $a\otimes a^\prime\in A\otimes A$. Since the 
de Rham differential $\dd^{\dR}: A \to \Omega^1_A$ 
is compatible with the right $H$-coactions on $A$ and on $\Omega^1_A$, 
those extend to a right $H$-coaction on the de Rham dg-algebra $\Omega^\bullet_A$.

\subsection{The Hamiltonian $G$-space $(T^\ast X,\omega,\mu)$}
In this subsection we provide an explicit description of the Hamiltonian
$G$-space that enters the symplectic reduction in \eqref{eqn:symplecticreduction}.
As before, let $X=\spec A$ be a smooth affine scheme together with an action $X\times G\to X$ 
of a smooth affine group scheme $G=\spec H$.
In this case the total space of the cotangent bundle $\pi : T^\ast X\to X$ is affine
too. It is concretely given by 
\begin{flalign}
T^\ast X \,=\, \spec \! \big(\Sym_A \T_{\! A}\big)\quad,
\end{flalign}
where $\Sym_A  \T_{\! A}$ denotes the symmetric algebra over $A$ of the $A$-module $\T_{\! A}$
of vector fields.
The projection map $\pi: T^\ast X\to X$ is algebraically described
by the algebra homomorphism $\pi^\ast : A\to \Sym_A \T_{\! A}$ that identifies
$A$ with the $0$-th symmetric power $A = \Sym^0_A \T_{\! A} \subseteq \Sym_A \T_{\! A}$.
The $H$-coactions on $A$ and $\T_{\! A}$ induce, by using again tensor product coactions, 
a right $H$-comodule algebra structure $\delta : \Sym_A\T_{\! A}\to \Sym_A\T_{\! A} \otimes H$ 
on the symmetric algebra. 
\sk

To describe the canonical symplectic structure on $T^\ast X$,  
we use that there is a short exact sequence of (relative) K\"ahler $1$-forms 
\begin{flalign}
\xymatrix{
0 \ar[r]~&~ \Sym_{A}\T_{\! A} \otimes_A^{} \Omega^1_A \ar[r]~&~ \Omega^1_{\Sym_A\T_{\! A}} \ar[r]~&~
\Omega^1_{\Sym_A\T_{\! A} /A}\ar[r]~&~0
}
\end{flalign}
associated with the algebra homomorphism $\pi^\ast : A\to \Sym_{A}\T_{\! A}$,
which in particular implies that every element in  $\Sym_{A}\T_{\! A} \otimes_A^{} \Omega^1_A$
may be regarded as a $1$-form on $\Sym_A\T_{\! A}$.  Applying the coevaluation map
$\mathrm{coev} : A\to \T_{\! A}\otimes_A^{} \Omega^1_A$ to the unit element 
$\oone\in A$ then allows us to define the {\em tautological $1$-form} 
\begin{flalign}\label{eqn:tautological}
\lambda \,:=\, \mathrm{coev}(\oone)\,\in\,\T_{\! A}\otimes_A^{} \Omega^1_A \, \subseteq\, 
\Sym_{A}\T_{\! A} \otimes_A^{} \Omega^1_A \,  \subseteq\,  \Omega^1_{\Sym_A\T_{\! A}}
\end{flalign}
on $T^\ast X$.
By construction, $\lambda$ is invariant under the $H$-coaction on $\Omega^1_{\Sym_A\T_{\! A}}$,
i.e.\ $\delta \lambda = \lambda\otimes \oone_H$.  The canonical symplectic form on $T^\ast X$
is obtained by taking the de Rham differential 
\begin{flalign}
\omega \,:=\, \dd^{\dR}\lambda\,\in\, \Omega^2_{\Sym_A\T_{\! A}}
\end{flalign}
of $\lambda$ and it is by construction invariant under the $H$-coaction too.
\sk

It remains to describe the moment map $\mu : T^\ast X \to \g^\ast$. Algebraically,
this is given by an algebra homomorphism $\mu^\ast : \Sym \,\g \to \Sym_A\T_{\! A}$
or equivalently, using that $\Sym\,\g$ is a free commutative algebra,  by
a linear map $\mu^\ast : \g \to \Sym_{A}\T_{\! A}$. We define
\begin{flalign}\label{eqn:momentmap}
\mu^\ast(\xi)\,:=\, - \iota_{\rho(\xi)}\lambda\,\in\,\T_{\! A}\, \subseteq \, \Sym_A\T_{\! A}\quad,
\end{flalign}
for all $\xi\in \g$, by contracting minus the tautological $1$-form \eqref{eqn:tautological}
with the vector fields obtained from the Lie algebra action $\rho : \g \to \T_{\! \Sym_A\T_{\! A}}$
associated with the $H$-coaction on $\Sym_A\T_{\! A}$. 
The map $\mu^\ast : \g \to \Sym_{A}\T_{\! A}$ is clearly equivariant with respect to
the adjoint $H$-coaction on $\g$ and it also satisfies the moment map condition
\begin{flalign}
\dd^{\dR}\mu^\ast(\xi)  =  - \big(\dd^{\dR} \iota_{\rho(\xi)} + \iota_{\rho(\xi)}\dd^{\dR}\big)\lambda + 
 \iota_{\rho(\xi)}\dd^{\dR}\lambda =-\mathcal{L}_{\rho(\xi)} \lambda +  \iota_{\rho(\xi)}\omega=   \iota_{\rho(\xi)}\omega\quad,
\end{flalign}
where we have used the definition $\omega = \dd^{\dR}\lambda$ of the canonical symplectic form
and also that the Lie derivative $\mathcal{L}_{\rho(\xi)}\lambda=0$ is zero as a consequence of 
invariance of $ \lambda$ under the $H$-coaction.

\subsection{\label{subsec:pushouts}The homotopy pushouts}
In this subsection we compute the two fiber products
of derived stacks in \eqref{eqn:mu-10} and \eqref{eqn:dCriteasydiagram}.
It is important to note that,  as a consequence of our hypotheses on $X$ and $G$, 
each derived stack in these diagrams is a derived affine scheme, i.e.\ it is of the form $\spec B$ for
some commutative dg-algebra in non-positive degrees $B\in \CDGA^{\leq 0}$,  which is
interpreted as the function dg-algebra of this derived stack. This implies that
we can dually compute these ($\infty$-categorical) 
fiber products in terms of {\em homotopy pushouts} in the model category $\CDGA^{\leq 0}$. 
We refer the reader to \cite{CDGA} for an excellent and very useful review of the standard
model structure on $\CDGA^{\leq 0}$. 
\sk

Let us start with the diagram \eqref{eqn:mu-10} that determines the derived stack 
$\mu^{-1}(0)$.  Passing over to the function dg-algebras,  we obtain the following
{\em homotopy} pushout diagram (indicated by `$h$') 
\begin{flalign}
\xymatrix@R=3em@C=3em{
\ar@{}[dr]^>{h} \ar[d]_-{\mu^\ast}\Sym\,\g\ar[r]^-{0^\ast}~&~ \bbK\ar@{-->}[d]\\
\Sym_A\T_{\! A} \ar@{-->}[r]_-{}~&~\O(\mu^{-1}(0))
}
\end{flalign}
in the model category $\CDGA^{\leq 0}$ that determines the commutative 
dg-algebra $\O(\mu^{-1}(0))\in\CDGA^{\leq 0}$ and hence the 
derived affine scheme $\mu^{-1}(0) =\spec\O(\mu^{-1}(0))$.
We compute this homotopy pushout via a standard technique from model category theory:
We replace the algebra homomorphism $0^\ast : \Sym\,\g \to \bbK$
that sends each generator $\xi\in\g$ to zero  by a weakly equivalent cofibration
$\widetilde{0}^\ast : \Sym\,\g \to \widetilde{\bbK}$ in $\CDGA^{\leq 0}$.
This allows us to compute  $\O(\mu^{-1}(0))\in\CDGA^{\leq 0}$ by 
the {\em ordinary} pushout
\begin{flalign}\label{eqn:mu-10pushout}
\xymatrix@R=3em@C=3em{
 \ar[d]_-{\mu^\ast}\Sym\,\g\ar[r]^-{\widetilde{0}^\ast}~&~ \widetilde{\bbK} \ar@{-->}[d]\\
\Sym_A\T_{\! A} \ar@{-->}[r]_-{}~&~\O(\mu^{-1}(0))
}
\end{flalign}
in $\CDGA^{\leq 0}$.  To obtain such a replacement,  let us introduce the
commutative dg-algebra
\begin{flalign}
\widetilde{\bbK}\,:=\, \Sym\Big(\g[1]\stackrel{\id}{\longrightarrow}\g\Big )\,\in\,\CDGA^{\leq 0}
\end{flalign}
that is given by the symmetric algebra of 
the acyclic cochain complex $\big(\g[1]\stackrel{\id}{\longrightarrow}\g\big )$ concentrated
in degrees $-1$ and $0$.  The algebra map $0^\ast : \Sym\,\g \to \bbK$
factors through the dg-algebra $\widetilde{\bbK}$ via
\begin{flalign}\label{eqn:factorization}
\xymatrix{
\Sym\,\g \ar[r]^-{\Sym \,i}~&~ \Sym\Big(\g[1]\stackrel{\id}{\longrightarrow}\g\Big ) \ar[r]^-{\Sym\, 0}~&~\Sym \,0 \, \cong\,  \bbK
}\quad,
\end{flalign}
where $i : \g \to \big(\g[1]\stackrel{\id}{\longrightarrow}\g\big )$ is the inclusion in degree $0$
and $0 : \big(\g[1]\stackrel{\id}{\longrightarrow}\g\big ) \to 0$ is the zero map.
The second map in \eqref{eqn:factorization} is a weak equivalence in $\CDGA^{\leq 0}$, i.e.\
a quasi-isomorphism on the underlying cochain complexes,  and the first map
is a semi-free extension and hence in particular a cofibration in $\CDGA^{\leq 0}$, cf.\ \cite{CDGA}.
This implies that we may use $\widetilde{0}^\ast := \Sym\,i : \Sym\,\g\to \widetilde{\bbK}$ 
to compute the pushout in \eqref{eqn:mu-10pushout}.  The resulting commutative dg-algebra
$\O(\mu^{-1}(0))\in\CDGA^{\leq 0}$ is then given explicitly by
\begin{flalign}\label{eqn:mu-10algebra}
\O(\mu^{-1}(0))\,=\, \Sym_A\T_{\! A} \otimes^{}_{\Sym\,\g} \widetilde{\bbK} \,\cong\, 
\Sym_A\Big( A\otimes \g[1] \stackrel{\mu^\ast}{\longrightarrow} \T_{\! A}\Big)\quad,
\end{flalign}
where with abuse of notation we denote by the same symbol
$\mu^\ast : A\otimes \g[1] \to \T_{\! A}\,,~a\otimes\xi \mapsto a\,\mu^\ast(\xi)$
the $A$-linear extension of the moment map $\mu^\ast : \g \to  \T_{\! A} \subseteq \Sym_A\T_{\! A}$ 
given in \eqref{eqn:momentmap}.
\sk

Let us consider now the diagram \eqref{eqn:dCriteasydiagram} 
that determines the derived stack $Z$.  Passing again 
over to the function dg-algebras,  we obtain the following
homotopy pushout diagram
\begin{flalign}\label{eqn:OZdiagramhomotopy}
\xymatrix@R=3em@C=3em{
\ar@{}[dr]^>{h} \ar[d]_-{(\dd^{\dR}f)^\ast}  \O(\mu^{-1}(0)) \ar[r]^-{0^\ast}~&~ A \ar@{-->}[d]\\
A \ar@{-->}[r]_-{}~&~\O(Z)
}
\end{flalign}
in the model category $\CDGA^{\leq 0}$ that determines the commutative 
dg-algebra $\O(Z)\in\CDGA^{\leq 0}$ and hence the 
derived affine scheme $ Z=\spec\O(Z)$. The two labeled dg-algebra homomorphisms in
this diagram are given concretely  by
\begin{subequations}
\begin{flalign}
0^\ast= \Sym_A 0 \,:\, \O(\mu^{-1}(0))\,\cong\, 
\Sym_A\Big( A\otimes \g[1] \stackrel{\mu^\ast}{\longrightarrow} \T_{\! A}\Big) ~\longrightarrow~\Sym_A 0\,\cong\,A\quad,
\end{flalign}
where $0: \big(A\otimes \g[1] \stackrel{\mu^\ast}{\longrightarrow} \T_{\! A}\big) \to 0$ 
denotes the zero map, and $(\dd^{\dR}f)^\ast : \O(\mu^{-1}(0))\to A$ is  defined on the 
generators of $\O(\mu^{-1}(0))$ by
\begin{flalign}
(\dd^{\dR}f)^\ast (a) = a\quad,\qquad (\dd^{\dR}f)^\ast (v) = \iota_{v}{\dd^\dR}f\quad,\qquad
(\dd^{\dR}f)^\ast (\xi)=0\quad,
\end{flalign}
\end{subequations}
for all $a\in A$, $v\in \T_{\! A}$ and $\xi\in\g[1]$. 
Note that the latter is indeed compatible with the differential on $\O(\mu^{-1}(0))$ 
(cf.\ \eqref{eqn:mu-10algebra}) because $\iota_{\mu^\ast(\xi)}\dd^{\dR}f = - \rho(\xi)(f) =0$, for all $\xi\in\g$, as a consequence of the 
invariance of $f\in A$ under the $H$-coaction $\delta:A\to A\otimes H$ on $A$.
\sk

Similarly to the computation of $\O(\mu^{-1}(0))$ from before,  
we compute the homotopy pushout \eqref{eqn:OZdiagramhomotopy} 
by replacing $0^\ast : \O(\mu^{-1}(0)) \to A$ by a weakly equivalent cofibration 
$\widetilde{0}^\ast : \O(\mu^{-1}(0))\to\widetilde{A}$ and then forming the ordinary pushout
\begin{flalign}\label{eqn:OZdiagram}
\xymatrix@R=3em@C=3em{
\ar[d]_-{(\dd^{\dR}f)^\ast}  \O(\mu^{-1}(0)) \ar[r]^-{\widetilde{0}^\ast}~&~ \widetilde{A} \ar@{-->}[d]\\
A \ar@{-->}[r]_-{}~&~\O(Z)
}
\end{flalign}
in $\CDGA^{\leq 0}$.  For this purpose we introduce the commutative dg-algebra
\begin{flalign}
\widetilde{A}\,:=\, \Sym_A\Big( \big(A\otimes \g[1] \stackrel{\mu^\ast}{\longrightarrow} \T_{\! A}\big) \otimes \big(\bbK[1] \stackrel{\id}{\longrightarrow}\bbK\big)\Big)\,\in\,\CDGA^{\leq 0}\quad,
\end{flalign}
where the middle $\otimes$ denotes the tensor product of cochain complexes,
and we observe that $0^\ast : \O(\mu^{-1}(0)) \to A$ factors as $\O(\mu^{-1}(0))\to\widetilde{A}\to A$
into a cofibration followed by a weak equivalence.  In more detail,
the latter factorization is given by applying the functor $\Sym_A$
to the maps
\begin{flalign}\label{eqn:TMPmaps}
\xymatrix{
\big(A\otimes \g[1] \stackrel{\mu^\ast}{\longrightarrow} \T_{\! A}\big) \otimes \bbK \ar[r]^-{\id\otimes i} 
~&~ \big(A\otimes \g[1] \stackrel{\mu^\ast}{\longrightarrow} \T_{\! A}\big) \otimes \big(\bbK[1] \stackrel{\id}{\longrightarrow}\bbK\big)
\ar[r]^-{0}~&~0
}
\end{flalign}
of cochain complexes of $A$-modules, where $i : \bbK \to \big(\bbK[1]\stackrel{\id}{\longrightarrow}\bbK\big )$ 
is the inclusion in degree $0$.  The second map is a weak equivalence between cofibrant objects (recall that $\T_{\! A}$
is projective), hence applying $\Sym_A$ defines a weak equivalence $\widetilde{A}\to A$.
Using again that $\T_{\! A}$ is projective, it follows that $\T_{\! A}$ is a retract
of a free $A$-module,  which implies that the first map in \eqref{eqn:TMPmaps}
is a retract of an injective map between cochain complexes of free $A$-modules.
As a consequence of this,  applying $\Sym_A$ to the first map in \eqref{eqn:TMPmaps}
defines a morphism $\O(\mu^{-1}(0))\to \widetilde{A}$ in $\CDGA^{\leq 0}$
that is a retract of a semi-free extension,  hence it is a cofibration, cf.\ \cite{CDGA}.
\sk

The commutative dg-algebra $\O(Z)\in\CDGA^{\leq 0}$ resulting from the pushout in \eqref{eqn:OZdiagram}
is then given explicitly by the graded algebra
\begin{subequations}\label{eqn:OZalgebra}
\begin{flalign}
\O(Z)\,=\, A\otimes_{\O(\mu^{-1}(0))}\widetilde{A} \,\cong\,\Sym_A\Big(\big(A\otimes \g[2]\big) \oplus  \T_{\! A}[1] \Big)
\end{flalign}
together with the differential defined on the generators by
\begin{flalign}
\dd a =0\quad,\qquad \dd v = \iota_v\dd^{\dR}f \quad,\qquad \dd \xi = \mu^\ast(\xi)\quad,
\end{flalign}
\end{subequations}
for all $a\in A$, $v\in \T_{\! A}[1]$ and $\xi\in \g[2]$.
Observe that $\O(Z)$ is canonically a right $H$-comodule dg-algebra
with respect to the tensor product coaction $\delta : \O(Z)\to\O(Z)\otimes H$ induced
from the given coactions on $A$ and $\T_{\! A}$ and the adjoint coaction on $\g$.


\section{\label{sec:functions}Function dg-algebra and quasi-coherent sheaves}
In the previous two sections we have obtained
an explicit model for the derived critical locus of a 
function $f:[X/G]\to \bbA_{\bbK}^1$ on the quotient stack 
of a smooth affine scheme $X=\spec A$ by the action of a smooth affine group scheme $G=\spec H$.
Our model is given by a derived quotient stack $\dCrit(f) \simeq [Z/G]$
for a derived affine scheme $Z=\spec\O(Z)$, whose
function dg-algebra $\O(Z)\in\CDGA^{\leq 0}$ we have spelled 
out concretely in \eqref{eqn:OZalgebra}. 
In this section we will describe
the function dg-algebra and the dg-category of quasi-coherent sheaves 
on the derived critical locus $\dCrit(f)$.  These algebraic objects are 
relevant for the study of quantization (see e.g.\ \cite{ToenICM}),
which we will not discuss in the present paper,  and they are also crucial for comparing
$\dCrit(f)$ with the usual BV formalism for Lie algebra actions in Section \ref{sec:comparison}.

\subsection{\label{subsec:functions}Function dg-algebra}
Let us start with the function dg-algebra $\O(\dCrit(f))$ of $\dCrit(f)\simeq[Z/G]$.
Using the general definition of function dg-algebras for derived stacks given in \cite{LoopSpaces},
see also \cite{SafronovLecture} for a review, we obtain that 
\begin{flalign}\label{eqn:functionsdCrit}
\O(\dCrit(f))\,=\,\holim\Big(
\xymatrix@C=1em{
\O(Z) \ar@<0.5ex>[r] \ar@<-0.5ex>[r] ~&~\O(Z)\otimes H  \ar@<1ex>[r] \ar[r] \ar@<-1ex>[r] ~&~\O(Z)\otimes H^{\otimes 2} 
\ar@<0.5ex>[r] \ar@<-0.5ex>[r]\ar@<1.5ex>[r] \ar@<-1.5ex>[r]~&~ \cdots
}
\Big)
\end{flalign}
is the homotopy limit in the model category of (possibly unbounded) dg-algebras
of the cosimplicial diagram associated with the right $H$-coaction $\delta : \O(Z)\to\O(Z)\otimes H$.
The latter can be computed by using the Dold-Kan correspondence
to pass from cosimplicial cochain complexes to double complexes,
followed by forming the $\prod$-total complex.  Spelling this out in more detail,
we obtain that the underlying cochain complex of the dg-algebra 
$\O(\dCrit(f))$ reads as
\begin{subequations}\label{eqn:OdCrit}
\begin{flalign}
\O(\dCrit(f))^k\,=\, \prod_{n+m=k} \O(Z)^{n}\otimes H^{+\otimes m}\quad,
\end{flalign}
for all $k\in\bbZ$,  where we recall that $H^+ = \ker(\epsilon: H\to \bbK)$ denotes the augmentation ideal.
The differential $\dd^\tot: \O(\dCrit(f))^k\to\O(\dCrit(f))^{k+1}$ 
is defined on elements $b\otimes \und{h} := b\otimes \Motimes_{i=1}^m h_i\in \O(\dCrit(f))$,  with 
$b\in\O(Z)^n$  homogeneous of degree $n$,  by
\begin{flalign}
\dd^\tot \big(b\otimes \und{h}\big)\,&=\, \dd b\otimes   \und{h} + (-1)^{n}\, \dd^{\mathbf{\Delta}} \big(b\otimes\und{h}\big) \quad,
\end{flalign}
where $\dd$ denotes the differential on $\O(Z)$
and $\dd^{\mathbf{\Delta}}$ is the cosimplicial differential defined by 
\begin{flalign}
\dd^{\mathbf{\Delta}} \big(b\otimes\und{h}\big) \,:=\, \delta(b)\otimes  \und{h} + \sum_{j=1}^{m} (-1)^j \, b\otimes 
\Delta_j( \und{h}) +(-1)^{m+1}\, b\otimes \und{h} \otimes \oone_H\quad, 
\end{flalign}
\end{subequations}
where $\delta : \O(Z)\to\O(Z)\otimes H$ is the coaction on $\O(Z)$, $\oone_H\in H$ is the unit element
and $\Delta_j(\und{h}) := h_1\otimes \cdots\otimes \Delta(h_j)\otimes \cdots \otimes h_m$ is
the application of the coproduct on the $j$-th factor.
Observe that $\O(\dCrit(f))= N^\bullet(G,\O(Z))$ 
is just the normalized group cohomology complex
of $G=\spec H$ with coefficients in the right $H$-comodule dg-algebra $\O(Z)$.
The product on $\O(\dCrit(f))$ is given by the cup-product from group cohomology,
which reads explicitly as
\begin{flalign}\label{eqn:OdCritproduct}
 \Big(b\otimes \Motimes_{i=1}^m h_i\Big)\,  \Big(b^\prime\otimes \Motimes_{j=1}^{m^\prime} h^\prime_j\Big)
 = (-1)^{n^\prime \, m}~b\,b^\prime_{\und{0}}\otimes \Motimes_{i=1}^m   h_i \, b^\prime_{\und{i}}\otimes  \Motimes_{j=1}^{m^\prime} h^\prime_j\quad,
\end{flalign}
where $n^\prime$ denotes the degree of $b^\prime\in\O(Z)$ and we have used the Sweedler notation 
$\delta^m(b^\prime) = b^\prime_{\und{0}}\otimes b^\prime_{\und{1}}\otimes \cdots\otimes b^\prime_{\und{m}}$
for the iterated application of the $H$-coaction. The unit element
of $\O(\dCrit(f))$ is given by $\oone\in\O(Z)$.  We would like to note
that our model for $\O(\dCrit(f))$ is {\em not} a strictly commutative dg-algebra, but rather
an $E_\infty$-algebra, i.e.\ a homotopy-coherently commutative dg-algebra.
The details about this $E_\infty$-algebra structure are not required in the present paper
and can be found e.g.\ in \cite{BergerFresse}. In what follows we shall regard
$\O(\dCrit(f))$ as an associative and unital dg-algebra with respect to the 
strictly associative product \eqref{eqn:OdCritproduct}.

\subsection{\label{subsec:QCoh}Quasi-coherent sheaf dg-category}
The function dg-algebra $\O(\dCrit(f))$ of the derived quotient stack $\dCrit(f) \simeq [Z/G]$
has the well-known deficit that it does not in general encode the derived stack faithfully.
A simple example that illustrates this phenomenon is as follows:
Let $G=\bbZ_2$ be the cyclic group of order $2$ and consider the
quotient stack $[\mathrm{pt}/\bbZ_2]$ determined by the trivial $\bbZ_2$-action on 
the point. (Note that the latter is a non-trivial stack, namely the classifying stack
of principal $\bbZ_2$-bundles.) Then the function dg-algebra $\O([\mathrm{pt}/\bbZ_2]) = N^\bullet(\bbZ_2,\bbK)$
is given by the normalized group cohomology complex of $\bbZ_2$ with coefficients in the trivial 
representation $\bbK$. For a field $\bbK$ of characteristic zero,  the latter is trivial
in non-zero degrees and one obtains a chain of weak equivalences
\begin{flalign}
\O([\mathrm{pt}/\bbZ_2]) \,=\, N^\bullet(\bbZ_2,\bbK)\, \simeq\, \bbK \,=\, \O(\mathrm{pt})
\end{flalign}
between the function dg-algebra of $[\mathrm{pt}/\bbZ_2]$ and that of the point $\mathrm{pt}$.
Hence, all information about the $\bbZ_2$-action in the quotient stack $[\mathrm{pt}/\bbZ_2]$
is lost when passing over to function dg-algebras.
\sk

To solve this issue,  one may assign to a derived stack a richer algebraic object,
namely its dg-category of quasi-coherent sheaves, see e.g.\ \cite{ToenICM}
and \cite{SafronovLecture} for rather non-technical introductions.
The latter can be interpreted roughly as the category of ``vector bundles'' over a given derived stack.
In our case of interest, which is given by a derived quotient stack $\dCrit(f)\simeq [Z/G]$
with $Z=\spec\O(Z)$ and $G=\spec H$ derived affines, the dg-category of quasi-coherent sheaves 
\begin{flalign}
\QCoh(\dCrit(f))\,\simeq\, \dgMod_{\O(Z)}^H
\end{flalign}
is the dg-category of (possibly unbounded) right $H$-$A_\infty$-comodule $\O(Z)$-dg-modules.
More explicitly,  an object in this dg-category is a (not necessarily bounded)
dg-module $M$ over the commutative dg-algebra $\O(Z)$
in \eqref{eqn:OZalgebra},  together with a suitably homotopy-coherent (called $A_\infty$)
generalization of a right $H$-coaction $\delta^M : M\to M\otimes H$  that is compatible with 
the $\O(Z)$-module structure on $M$ and with the $H$-coaction $\delta : \O(Z)\to\O(Z)\otimes H$ on $\O(Z)$.
The details for such $A_\infty$-comodules 
are not required in the present paper and can be found e.g.\ in \cite{dgCat}.
\sk

It is important to emphasize that from this richer quasi-coherent sheaf perspective,  
one can recover the function dg-algebra $\O(\dCrit(f))$ via the following construction: 
The right $H$-comodule dg-algebra $\O(Z)$ may be regarded as a (strict) right $H$-$A_\infty$-comodule
$\O(Z)$-dg-module, i.e.\ it defines an object in the dg-category $\QCoh(\dCrit(f))$.
(This object is the monoidal unit for the canonical symmetric monoidal structure
on this dg-category.) Taking the endomorphisms of any object in a dg-category
defines a dg-algebra and for the object $\O(Z)\in \QCoh(\dCrit(f))$ one finds that
\begin{flalign}
\und{\End}(\O(Z))\,\simeq\, \O(\dCrit(f))
\end{flalign}
is equivalent to the function dg-algebra of $\dCrit(f)$.


\section{\label{sec:symplectic}$(-1)$-shifted symplectic structure}
Using our concrete model for $\dCrit(f) \simeq [Z/G]$,  we can directly 
write down a simple expression for a $(-1)$-shifted symplectic structure.  
We will then confirm in Subsection \ref{subsec:symplcompare} that this agrees 
with the canonical, but more abstract, definition
of the $(-1)$-shifted symplectic structure on $\dCrit(f)$ via Lagrangian intersections,
which we have briefly recalled in the last paragraph of Section \ref{sec:DAG}.

\subsection{Explicit formula}
In analogy to  the function dg-algebra in \eqref{eqn:functionsdCrit},
the differential forms on the derived quotient stack $\dCrit(f) \simeq [Z/G]$
may be computed by the homotopy limit
\begin{flalign}
\Omega^\bullet(\dCrit(f))\,=\,\holim\Big(
\xymatrix@C=1em{
\Omega^\bullet_{\O(Z)} \ar@<0.5ex>[r] \ar@<-0.5ex>[r] ~&~\Omega^{\bullet}_{\O(Z)\otimes H}  \ar@<1ex>[r] \ar[r] \ar@<-1ex>[r] ~&~
\Omega^\bullet_{\O(Z)\otimes H^{\otimes 2}} 
\ar@<0.5ex>[r] \ar@<-0.5ex>[r]\ar@<1.5ex>[r] \ar@<-1.5ex>[r]~&~ \cdots
}
\Big)\quad.
\end{flalign}
More concretely,  a model for the cochain complex $\Omega^p(\dCrit(f))$ of $p$-forms on $\dCrit(f)$,
for $p\in\bbZ_{\geq 0}$, is given by the $\prod$-total complex of the double complex
\begin{flalign}\label{eqn:doublecomplex}
\xymatrix{
(\Omega^p_{\O(Z)})^{0} \ar[r]^-{\dd^{\mathbf{\Delta}}}~&~(\Omega^p_{\O(Z)\otimes H})^{0}  \ar[r]^-{\dd^{\mathbf{\Delta}}}
~&~ (\Omega^p_{\O(Z)\otimes H^{\otimes 2}})^{0}  \ar[r]^-{\dd^{\mathbf{\Delta}}}~&~\cdots\\
\ar[u]^-{\dd}(\Omega^p_{\O(Z)})^{-1} \ar[r]^-{\dd^{\mathbf{\Delta}}}~&~\ar[u]^-{\dd}(\Omega^p_{\O(Z)\otimes H})^{-1}  \ar[r]^-{\dd^{\mathbf{\Delta}}}
~&~\ar[u]^-{\dd}(\Omega^p_{\O(Z)\otimes H^{\otimes 2}})^{-1}  \ar[r]^-{\dd^{\mathbf{\Delta}}}~&~\cdots\\
\ar[u]^-{\dd}(\Omega^p_{\O(Z)})^{-2} \ar[r]^-{\dd^{\mathbf{\Delta}}}~&~\ar[u]^-{\dd}(\Omega^p_{\O(Z)\otimes H})^{-2}  \ar[r]^-{\dd^{\mathbf{\Delta}}}
~&~\ar[u]^-{\dd}(\Omega^p_{\O(Z)\otimes H^{\otimes 2}})^{-2}  \ar[r]^-{\dd^{\mathbf{\Delta}}}~&~\cdots\\
\ar[u]^-{\dd} \vdots~&~\ar[u]^-{\dd} \vdots~&~\ar[u]^-{\dd} \vdots~&~\ddots
}
\end{flalign}
where the vertical degrees (denoted as superscripts) are induced by $\O(Z)$ and
the differentials $\dd$ and $\dd^{\mathbf{\Delta}}$ 
are given by the extension to differential forms of the differentials in \eqref{eqn:OdCrit}.
Recalling the commutative dg-algebra $\O(Z)$ from \eqref{eqn:OZalgebra},
we observe that,  by using the coevaluation map $\mathrm{coev} : A\to \T_{\! A} \otimes_A \Omega^1_A$,
one can define the $1$-form
\begin{flalign}\label{eqn:lambdaTastX-1}
\lambda_{T^\ast[-1] X} \,:=\, \mathrm{coev}(\oone)\,\in\, \T_{\! A}[1] \otimes_A \Omega^1_A \,\subseteq (\Omega^1_{\O(Z)})^{-1}
\end{flalign}
in vertical degree $-1$ and horizontal degree $0$.
Furthermore, using the coevaluation map $\mathrm{coev} : \bbK \to \g \otimes \g^\ast $ for the Lie algebra $\g$
and its dual $\g^\ast = H^+/H^{+ 2}$,  together with the linear map $j : \g^\ast \to \Omega^1_H\, ,~[\theta]\mapsto
(\dd^{\dR}\theta_{\und{1}})\, S(\theta_{\und{2}})$ that assigns right $H$-invariant $1$-forms on $H$,
one can define the $1$-form
\begin{flalign}\label{eqn:lambdaBG-1}
\lambda_{T^\ast[-1]BG} \,:=\, (\id\otimes j)\mathrm{coev}(1)\,\in\, \g[2]\otimes \Omega^1_H \,\subseteq\, (\Omega^1_{\O(Z)\otimes H})^{-2}
\end{flalign}
in vertical degree $-2$ and horizontal degree $1$. 
One can show by an explicit calculation that the two elements $\lambda_{T^\ast[-1] X}$ and $\lambda_{T^\ast[-1]BG}$ 
of the double  complex \eqref{eqn:doublecomplex} satisfy the three conditions
\begin{flalign}
\dd\lambda_{T^\ast[-1] X} \,=\,\dd^{\dR}f~~,\quad \dd^{\mathbf{\Delta}}\lambda_{T^\ast[-1] X} + \dd\lambda_{T^\ast[-1]BG} =\,0\,~~,\quad
\dd^{\mathbf{\Delta}}\lambda_{T^\ast[-1]BG} \,=\,0\quad.
\end{flalign}
As a consequence,  setting
\begin{flalign}\label{eqn:shiftedsymplectic}
\omega_{\dCrit(f)}\,:=\, \dd^{\dR} \big( \lambda_{T^\ast[-1] X} - \lambda_{T^\ast[-1]BG}\big)\,\in\,\Omega^2(\dCrit(f))
\end{flalign}
defines a $\dd^{\dR}$-exact $2$-form on $\dCrit(f)$ 
that is further a $(-1)$-cocycle in the $\prod$-total complex $\Omega^2(\dCrit(f))$,
i.e.\ $\dd^{\tot}\omega_{\dCrit(f)}=0$ for the total differential defined by $\dd$ and $\dd^{\mathbf{\Delta}}$.
The $2$-form \eqref{eqn:shiftedsymplectic} is non-degenerate, hence
it defines a $(-1)$-shifted symplectic structure on $\dCrit(f)$.

\subsection{\label{subsec:symplcompare}Relationship to the canonical $(-1)$-shifted symplectic structure}
Our $(-1)$-shifted symplectic structure in \eqref{eqn:shiftedsymplectic} agrees
with the canonical one that is obtained from the theory of derived intersections of (shifted) Lagrangians, see
e.g.\ \cite{Albin} and references therein. This more abstract point of view is particularly useful to explain
the relative sign in \eqref{eqn:shiftedsymplectic}. We will now briefly explain how the relevant computations
can be performed.
\sk

Consider first the fiber product of derived stacks
\begin{flalign}\label{eqn:intersection1}
\xymatrix@R=3em@C=3em{
\ar@{-->}[d] [T^\ast X/\!\! /G] \ar@{-->}[r]~&~ [\mathrm{pt}/G] \ar[d]^-{0}\\
[T^\ast X/G] \ar[r]_-{\mu}~&~ [\g^\ast /G]
}
\end{flalign}
that determines the symplectic reduction $[T^\ast X/\!\! /G]$
and hence the total space of the cotangent bundle $T^\ast[X/G]\simeq [T^\ast X/\!\! /G]$.
The derived stack $[\g^\ast /G]$ in the bottom right corner 
carries a canonical $1$-shifted symplectic structure because
$[\g^\ast /G] \simeq T^\ast[1]BG$  is equivalent to the total space
of the $1$-shifted cotangent bundle over $BG := [\mathrm{pt}/G]$.
This shifted symplectic structure is given by applying $\dd^\dR$
on the tautological $1$-form (of degree $1$)
$\lambda_{T^\ast[1]BG}\in \Omega^1([\g^\ast/G])$,
which explicitly takes the same form as its shifted analog in
\eqref{eqn:lambdaBG-1}.  Pulling back this form along $\mu$ yields
\begin{flalign}
\mu^\ast \big(\lambda_{T^\ast[1]BG }\big)\,=\, - \dd^{\tot} \lambda_{T^\ast X}\quad,
\end{flalign}
where $\lambda_{T^\ast X} \in \Omega^1([T^\ast X/G])$ 
is the tautological $1$-form (of degree $0$) on $T^\ast X$, cf.\ \eqref{eqn:tautological}.
Hence, $\mu$ is a shifted Lagrangian with Lagrangian structure $-\dd^\dR \lambda_{T^\ast X}$.
Replacing as in Subsection \ref{subsec:pushouts} the zero map $0$ in \eqref{eqn:intersection1} 
by the weakly equivalent fibration $\widetilde{0}:\widetilde{\mathrm{pt}}:=\spec\widetilde{\bbK} \to\g^\ast$ 
given in \eqref{eqn:factorization}, we obtain for the pullback
\begin{flalign}
\widetilde{0}^\ast\big(\lambda_{T^\ast[1]BG }\big)\,=\, \dd^{\tot} \lambda_{T^\ast BG }\quad, 
\end{flalign}
where $\lambda_{T^\ast BG} \in \Omega^1([\widetilde{\mathrm{pt}}/G])$
is of degree $0$ and defined analogously to \eqref{eqn:lambdaBG-1}. 
Hence, $\widetilde{0}$ is a shifted Lagrangian with Lagrangian structure $ \dd^{\dR} \lambda_{T^\ast BG }$.
The canonical $0$-shifted symplectic structure on the fiber product of shifted Lagrangians 
in \eqref{eqn:intersection1} is then given by the $0$-cocycle 
\begin{flalign}\label{eqn:canonicalformsymplecticreduction}
\omega_{[T^\ast X/\!\! /G]} \,=\, \dd^{\dR}\big(  \lambda_{T^\ast X} +  \lambda_{T^\ast BG }\big)\,\in\,
\Omega^2\big([T^\ast X/\!\! /G]\big)\quad.
\end{flalign}

Let us consider now the  fiber product of derived stacks
\begin{flalign}\label{eqn:intersection2}
\xymatrix@R=3em@C=3em{
\ar@{-->}[d] \dCrit(f) \ar@{-->}[r]~&~ [X/G] \ar[d]^-{0}\\
[X/G] \ar[r]_-{\dd^\dR f}~&~ [T^\ast X/\!\! /G]\simeq T^\ast[X/G]
}
\end{flalign}
that determines the derived critical locus. Pulling back
\eqref{eqn:canonicalformsymplecticreduction} along $\dd^\dR f$ yields
\begin{flalign}
(\dd^\dR f)^\ast\big( \omega_{[T^\ast X/\!\! /G]} \big)\,=\, \dd^\dR \,\dd^\dR f \,=\,0\quad,
\end{flalign}
hence the section $\dd^\dR f$ is Lagrangian with a trivial Lagrangian structure.
Replacing as in Subsection \ref{subsec:pushouts} the zero section $0$ in \eqref{eqn:intersection2} 
by the weakly equivalent fibration $\widetilde{0}:\widetilde{X}:=\spec\widetilde{A} \to T^\ast X$ 
given in \eqref{eqn:TMPmaps}, we obtain for the pullback
\begin{flalign}
\widetilde{0}^\ast\big(  \omega_{[T^\ast X/\!\! /G]}  \big) \,=\, \dd^\tot\dd^\dR \big(
 \lambda_{T^\ast[-1] X} -\lambda_{T^\ast[-1] BG } \big)\quad,
\end{flalign}
hence $\widetilde{0}$ is a Lagrangian with Lagrangian structure 
$\dd^\dR \big( \lambda_{T^\ast[-1] X} -\lambda_{T^\ast[-1] BG } \big)$.
(The two summands were defined explicitly in \eqref{eqn:lambdaTastX-1} and \eqref{eqn:lambdaBG-1}.)
The resulting canonical $(-1)$-shifted symplectic structure on $\dCrit(f)$
is then given by our simple formula \eqref{eqn:shiftedsymplectic} from the previous subsection.


\section{\label{sec:comparison}Comparison to the BV formalism for Lie algebra actions}
Let us start by recalling that our explicit model for
the derived critical locus of a function $f : [X/G]\to \bbA^1_\bbK$
on a quotient stack is given by a derived quotient stack $\dCrit(f)\simeq [Z/G]$,
where $Z=\spec\O(Z)$ is a derived affine scheme whose dg-algebra
of functions $\O(Z)\in \CDGA^{\leq 0}$ is given in \eqref{eqn:OZalgebra}.
To relate this to the ordinary BV formalism for Lie algebra
actions,  we consider the $\g$-action on the dg-algebra 
$\O(Z)$ that is induced via \eqref{eqn:Liealgebraaction}
from the right $H$-coaction $\delta : \O(Z)\to\O(Z)\otimes H$.
This allows us to define the formal stack
\begin{flalign}
\BV(f) \,:= \,[Z/\g]\quad,
\end{flalign}
as the quotient of $Z$ by the formal group integrating $\g$ \cite{SafronovPoissonLie}. It is the analog for infinitesimal symmetries of the derived quotient stack
$\dCrit(f)\simeq [Z/G]$ that models the derived critical locus.
The dg-algebra of functions of $\BV(f)$ is given by the Chevalley-Eilenberg cochains
\begin{flalign}
\O(\BV(f))\,=\, \CE^\bullet(\g,\O(Z))
\end{flalign}
of the Lie algebra $\g$ with coefficients in the representation $\O(Z)$.
More explicitly,  the underlying cochain complex of the dg-algebra 
$\O(\BV(f))$ reads as
\begin{subequations}
\begin{flalign}
\O(\BV(f))^k \,=\, \bigoplus_{n+m=k} \O(Z)^n\otimes{\textstyle \bigwedge^m} \g^\ast\,\cong\, \bigoplus_{n+m=k} \Hom^{}_\bbK\big({\textstyle \bigwedge^m} \g , \O(Z)^n\big)\quad,
\end{flalign}
for all $k\in\bbZ$, and the differential $\dd^\tot: \O(\BV(f))^k\to \O(\BV(f))^{k+1}$
is defined on homogeneous elements $(\varphi : \bigwedge^m\g\to \O(Z)^n) \in \O(\BV(f))$ by
\begin{flalign}
\dd^\tot\varphi \,&=\, \dd \varphi + (-1)^{n}\, \dd^{\CE} \varphi\quad,
\end{flalign}
where $\dd$ denotes the differential on $\O(Z)$
and $\dd^{\CE}$ is the Chevalley-Eilenberg differential. 
Recall that the latter is given explicitly  by
\begin{multline}
(\dd^{\CE}\varphi)(\xi_1,\dots,\xi_{m+1})\,=\, \sum_{i}
(-1)^{i+1}\, \rho(\xi_i)\big(\varphi(\xi_1,\dots,\widehat{\xi}_i,\dots, \xi_{m+1})\big)\\
+\sum_{i<j}(-1)^{i+j}\,  \varphi\big([\xi_i,\xi_j],\xi_1,\dots,\widehat{\xi}_i,\dots,\widehat{\xi}_j,\dots,\xi_{m+1}\big)\quad,
\end{multline}
\end{subequations}
for all $\xi_1,\dots,\xi_{m+1}\in \g$.
The product of the dg-algebra $\O(\BV(f))$ is defined on 
homogeneous elements $\varphi : \bigwedge^m\g\to \O(Z)^n$
and $\varphi^\prime:\bigwedge^{m^\prime}\g\to \O(Z)^{n^\prime}$ by
\begin{flalign}
(\varphi\, \varphi^\prime)(\xi_1,\dots,\xi_{m+m^\prime})
= \frac{(-1)^{n^\prime\, m}}{m!m^\prime !}\!\!\! \sum_{\sigma\in\Sigma_{m+m^\prime}}\!\!\!
\mathrm{sgn}(\sigma)\, \varphi(\xi_{\sigma(1)},\dots,\xi_{\sigma(m)})\,\varphi^\prime(\xi_{\sigma(m+1)},\dots,\xi_{\sigma(m+m^\prime)})\quad,
\end{flalign}
for all $\xi_1,\dots,\xi_{m+m^\prime}\in \g$, where $\Sigma_{m+m^\prime}$ denotes
the permutation group on $m+m^\prime$ letters and $\mathrm{sgn}(\sigma)$ the sign of a permutation $\sigma$.
The unit element of  $\O(\BV(f))$ is given by $\oone \in\O(Z)$.
Note that the dg-algebra $\O(\BV(f))$ is precisely the one of 
the BV formalism: The generators in non-positive degrees, which coincide with
the generators of $\O(Z)$ (cf.\ \eqref{eqn:OZalgebra}),  describe
(observables for) gauge fields $a\in A$,  antifields $v\in \T_{\! A}[1]$ and antifields for ghosts
$\xi\in \g[2]$. The Chevalley-Eilenberg generators $\theta\in\g^\ast$ 
are of total degree $1$ and 
describe (observables for) ghost fields. The total differential
$\dd^{\tot}$ on $\O(\BV(f))$ is the BV differential.
\sk

The dg-algebra $\O(\BV(f))$ of the usual BV formalism can be related directly
to our function dg-algebra $\O(\dCrit(f))$ of the derived critical locus (cf.\ \eqref{eqn:OdCrit}).
The relevant construction is given by the {\em van Est map}
from Lie groupoid to Lie algebroid cohomology, see e.g.\ \cite{Weinstein,Mehta}.
For our example,  the van Est map is the cochain map
\begin{subequations}\label{eqn:vEmap}
\begin{flalign}
\mathrm{vE} \,:\, \O(\dCrit(f))~\longrightarrow~\O(\BV(f))
\end{flalign}
that is defined on homogeneous elements $b\otimes \und{h} := b\otimes \Motimes_{i=1}^m h_i\in \O(\dCrit(f))$ by
\begin{flalign}
\mathrm{vE}\big(b\otimes \und{h} \big)(\xi_1,\dots,\xi_m) 
\,:=\, b \sum_{\sigma\in\Sigma_{m}} \mathrm{sgn}(\sigma)\, \xi_{\sigma(1)}(h_1)\,\cdots\, \xi_{\sigma(m)}(h_m)\quad,
\end{flalign}
\end{subequations}
for all $\xi_1,\dots,\xi_m\in\g$, where the expressions $\xi_{\sigma(i)}(h_i)$ are defined by
evaluating the relative derivations $\xi_{\sigma(i)}\in \g\cong \Der_\epsilon(H,\bbK)$ on $h_i\in H$.
(Note that,  due to antisymmetrization, 
this map is zero for all $m > \dim(\g)$. Hence, it is defined on the product \eqref{eqn:OdCrit}.)
The van Est map is a morphism of associative and unital dg-algebras,  see e.g.\ \cite[Proposition 6.2.3]{Mehta},
and we believe that it can be upgraded to an $\infty$-morphism between $E_\infty$-algebras,
although we are not aware of a formal proof of this statement.
\sk

We conclude by showing that the canonical
$(-1)$-shifted symplectic structure \eqref{eqn:shiftedsymplectic} on $\dCrit(f)$
corresponds via the van Est map to the usual $(-1)$-shifted symplectic structure on $\BV(f)$ 
(whose dual is the antibracket) of the BV formalism.
To obtain an extension of the van Est map \eqref{eqn:vEmap}
to differential forms, it is convenient to 
pass through the equivalent Cartan-Getzler model
$\Omega^\bullet(\dCrit(f))\simeq \Omega^\bullet_{\mathrm{CG}}([Z/G])$
for differential forms on a derived quotient stack, see
e.g.\ \cite[Section 2]{Yeung} for a fully explicit description. 
Then the following van Est map
\begin{flalign}\label{eqn:vEforms}
\xymatrix{
\Omega^p_{\mathrm{CG}}([Z/G])
=N^\bullet\big(G,\Omega^p_{\mathrm{Car}}(\g^\ast,\O(Z))\big)\ar[r]^-{\mathrm{vE}}
~&~ \CE^\bullet\big(\g,\Omega^p_{\mathrm{Car}}(\g^\ast,\O(Z))\big)=:\Omega^p([Z/\g])
}
\end{flalign}
yields a comparison map 
between differential forms on $\dCrit(f)\simeq [Z/G]$ and
$\BV(f)=[Z/\g]$. In the Cartan-Getzler model,
the canonical $(-1)$-shifted symplectic structure  \eqref{eqn:shiftedsymplectic}
takes a very similar form. In more detail, we have that 
\begin{subequations}\label{eqn:omegaCG}
\begin{flalign}
\omega_{\dCrit(f)}^{\mathrm{CG}}\,=\, 
\dd^{\dR} \big( \lambda^{\mathrm{CG}}_{T^\ast[-1] X} - \lambda^{\mathrm{CG}}_{T^\ast[-1]BG}\big)\,\in\,\Omega^2_{\mathrm{CG}}([Z/G])\quad,
\end{flalign}
where both
\begin{flalign}
\lambda^{\mathrm{CG}}_{T^\ast[-1] X} \,&:=\,
\mathrm{coev}(\oone)\in \T_{\! A}[1]\otimes_A\Omega^1_A\, \subseteq\,
\Omega^1_{\O(Z)} \,\subseteq\, \Omega^1_{\mathrm{Car}}(\g^\ast,\O(Z))
\end{flalign}
and
\begin{flalign}
\lambda^{\mathrm{CG}}_{T^\ast[-1]BG} \,&:=\, \mathrm{coev}(1)\in \g[2]\otimes\g^\ast\, \subseteq\,
\O(Z)\otimes \g^\ast \, \subseteq \, \Omega^1_{\mathrm{Car}}(\g^\ast,\O(Z))
\end{flalign}
\end{subequations}
are defined through $G$-invariant Cartan $1$-forms. As a consequence,
$\omega_{\dCrit(f)}^{\mathrm{CG}}$ is represented by a $G$-invariant Cartan $2$-form,
on which the van Est map \eqref{eqn:vEforms} acts as the identity.
The resulting $(-1)$-shifted symplectic structure 
$\mathrm{vE}(\omega_{\dCrit(f)}^{\mathrm{CG}})\in \Omega^2([Z/\g])$ on $\BV(f)$ 
is the usual shifted symplectic structure from the BV formalism, 
see e.g.\ \cite{CMR}. Indeed, the two contributions to 
$\mathrm{vE}(\omega_{\dCrit(f)}^{\mathrm{CG}})$ coming from 
\eqref{eqn:omegaCG} respectively pair antifields to fields 
and antifields for ghosts to ghosts. 


\section*{Acknowledgments}
We thank Damien Calaque for sharing with us, 
briefly after the present paper appeared on the arXiv, 
a draft for \cite{AnelCalaque}, 
which contains closely related results obtained independently. 
We also would like to thank the anonymous referee 
for their valuable comments that helped us to improve the paper. 
A.S.\ gratefully acknowledges the financial support of 
the Royal Society (UK) through a Royal Society University 
Research Fellowship (UF150099), a Research Grant (RG160517) 
and two Enhancement Awards (RGF\textbackslash EA\textbackslash 180270 and RGF\textbackslash EA\textbackslash 201051).


\end{document}